\documentclass[longauth]{aa} 

\usepackage{graphicx}
\usepackage{txfonts}
\usepackage{amsmath,amsfonts,amssymb}
\usepackage{booktabs}
\usepackage{textcomp,gensymb}
\usepackage{changes}
\usepackage{academicons}
\usepackage{natbib,twoopt}
\usepackage[breaklinks=true]{hyperref} 
\usepackage{array,booktabs,ragged2e}
\usepackage{soul}
\newcolumntype{R}[1]{>{\RaggedLeft\arraybackslash}p{#1}}
\bibpunct{(}{)}{;}{a}{}{,} 
\makeatletter
\newcommandtwoopt{\citeads}[3][][]{\href{http://adsabs.harvard.edu/abs/#3}%
{\def\hyper@linkstart##1##2{}%
\let\hyper@linkend\@empty\citealp[#1][#2]{#3}}}
\newcommandtwoopt{\citepads}[3][][]{\href{http://adsabs.harvard.edu/abs/#3}%
{\def\hyper@linkstart##1##2{}%
\let\hyper@linkend\@empty\citep[#1][#2]{#3}}}
\newcommandtwoopt{\citetads}[3][][]{\href{http://adsabs.harvard.edu/abs/#3}%
{\def\hyper@linkstart##1##2{}%
\let\hyper@linkend\@empty\citet[#1][#2]{#3}}}
\newcommandtwoopt{\citeyearads}[3][][]%
{\href{http://adsabs.harvard.edu/abs/#3}
{\def\hyper@linkstart##1##2{}%
\let\hyper@linkend\@empty\citeyear[#1][#2]{#3}}}
\makeatother
\newcommand{\sophi}{SO/PHI}
\newcommand{\hmi}{HMI}

\newcommand{\hrt}{SO/PHI-HRT}

\newcommand{\bm}[1]{\mbox{\boldmath$#1$\unboldmath}}

\makeatletter
\renewcommand*\aa@pageof{, page \thepage{} of \pageref*{LastPage}}
\begin{document}

\title{Magnetic fields inferred by Solar Orbiter: A comparison between SO/PHI-HRT and SDO/HMI}


   \author{
J.~Sinjan\inst{1}\orcid{0000-0002-5387-636X}\thanks{\hbox{Corresponding author: J. Sinjan} \hbox{\email{sinjan@mps.mpg.de}}}
     \and
    D.~Calchetti\inst{1}\orcid{0000-0003-2755-5295} \and 
    J.~Hirzberger\inst{1} \and
    F. Kahil\inst{1}\orcid{0000-0002-4796-9527} \and
    G.~Valori\inst{1}\orcid{0000-0001-7809-0067} \and
    S.K.~Solanki\inst{1}\orcid{0000-0002-3418-8449} \and
    K.~Albert\inst{1}\orcid{0000-0002-3776-9548} \and
    N. Albelo~Jorge\inst{1} \and
    A.~Alvarez-Herrero\inst{2}\orcid{0000-0001-9228-3412} \and
    T.~Appourchaux\inst{3}\orcid{0000-0002-1790-1951} \and 
    L.R.~Bellot~Rubio\inst{4}\orcid{0000-0001-8669-8857} \and
    J.~Blanco~Rodr\'\i guez\inst{5}\orcid{0000-0001-9228-3412} \and
    A.~Feller\inst{1} \and
    A.~Gandorfer\inst{1}\orcid{0000-0002-9972-9840} \and
    D.~Germerott\inst{1} \and
    L.~Gizon\inst{1,10}\orcid{0000-0001-7696-8665} \and
    J.M.~Gómez~Cama\inst{7}\orcid{0000-0003-0173-5888} \and
    L.~Guerrero\inst{1} \and
    P.~Gutierrez-Marques\inst{1}\orcid{0000-0003-2797-0392} \and
    M.~Kolleck\inst{1} \and
    A.~Korpi-Lagg\inst{1}\orcid{0000-0003-1459-7074} \and
    H.~Michalik\inst{6} \and
    A.~Moreno~Vacas\inst{4}\orcid{0000-0002-7336-0926} \and
    D.~Orozco~Su\' arez\inst{4}\orcid{0000-0001-8829-1938} \and
    I.~P\' erez-Grande\inst{8}\orcid{0000-0002-7145-2835} \and 
    E.~Sanchis~Kilders\inst{5}\orcid{0000-0002-4208-3575} \and
    M.~Balaguer~Jiménez\inst{4}\orcid{0000-0003-4738-7727}\and
    J.~Schou\inst{1}\orcid{0000-0002-2391-6156} \and
    U.~Sch\" uhle\inst{1}\orcid{0000-0001-6060-9078} \and
    J.~Staub\inst{1}\orcid{0000-0001-9358-5834} \and
    H.~Strecker\inst{4}\orcid{0000-0003-1483-4535} \and
    J.C.~del~Toro~Iniesta\inst{4}\orcid{0000-0002-3387-026X} \and
    R.~Volkmer\inst{9} \and
    J.~Woch\inst{1}\orcid{0000-0001-5833-3738}
    }

   \institute{
         Max-Planck-Institut f\"ur Sonnensystemforschung, Justus-von-Liebig-Weg 3,
         37077 G\"ottingen, Germany \\ \email{solanki@mps.mpg.de}
         \and
         Instituto Nacional de T\' ecnica Aeroespacial, Carretera de
         Ajalvir, km 4, E-28850 Torrej\' on de Ardoz, Spain
         \and
         Univ. Paris-Sud, Institut d’Astrophysique Spatiale, UMR 8617,
         CNRS, B\^ atiment 121, 91405 Orsay Cedex, France
         \and
         Instituto de Astrofísica de Andalucía (IAA-CSIC), Apartado de Correos 3004,
         E-18080 Granada, Spain \\ \email{jti@iaa.es}
         \and
         Universitat de Val\`encia, Catedr\'atico Jos\'e Beltr\'an 2, E-46980 Paterna-Valencia, Spain
         \and
         Institut f\"ur Datentechnik und Kommunikationsnetze der TU
         Braunschweig, Hans-Sommer-Str. 66, 38106 Braunschweig,
         Germany
         \and
         University of Barcelona, Department of Electronics, Carrer de Mart\'\i\ i Franqu\`es, 1 - 11, 08028 Barcelona, Spain
         \and
         Instituto Universitario "Ignacio da Riva", Universidad Polit\'ecnica de Madrid, IDR/UPM, Plaza Cardenal Cisneros 3, E-28040 Madrid, Spain
         \and
         Leibniz-Institut für Sonnenphysik, Sch\" oneckstr. 6, D-79104 Freiburg, Germany
         \and
         Institut f\"ur Astrophysik, Georg-August-Universit\"at G\"ottingen, Friedrich-Hund-Platz 1, 37077 G\"ottingen, Germany}

   \date{Received Dec XX, 2022; accepted, Mar XX, 2023}

 
  \abstract
   {The High Resolution Telescope (HRT) of the Polarimetric and Helioseismic Imager on board the Solar Orbiter spacecraft (\sophi) and the Helioseismic and Magnetic Imager (HMI) on board the Solar Dynamics Observatory (SDO) both infer the photospheric magnetic field from polarised light images. \sophi\ is the first magnetograph to move out of the Sun--Earth line and will provide unprecedented access to the Sun's poles. This provides excellent opportunities for new research wherein the magnetic field maps from both instruments are used simultaneously.}
   {We aim to compare the magnetic field maps from these two instruments and discuss any possible differences between them.}
   {We used data from both instruments obtained during Solar Orbiter's inferior conjunction on 7 March 2022. The HRT data were additionally treated for geometric distortion and degraded to the same resolution as HMI. The HMI data were re-projected to correct for the $3^\circ$ separation between the two observatories.}
   {\hrt\ and \hmi\ produce remarkably similar line-of-sight magnetograms, with a slope coefficient of $0.97$, an offset below $1\,$G, and a Pearson correlation coefficient of $0.97$. However, \hrt\ infers weaker line-of-sight fields for the strongest fields. As for the vector magnetic field, \hrt\ was compared to both the $720$-second and $90$-second \hmi\ vector magnetic field: \hrt\ has a closer alignment with the $90$-second \hmi\ vector. In the weak signal regime ($< 600$\,G), \hrt\ measures stronger and more horizontal fields than \hmi,\ very likely due to the greater noise in the \hrt\ data. In the strong field regime  ($\gtrsim 600$\,G), HRT infers lower field strengths but with similar inclinations (a slope of $0.92$) and azimuths (a slope of $1.02$). The slope values are from the comparison with the \hmi\ $90$-second vector. Possible reasons for the differences found between \hrt\ and \hmi\ magnetic field parameters are discussed.}
   {}


   \maketitle
%

\section{Introduction}
The Solar Orbiter \citep[see][]{mueller2013solar, muller2020solar} spacecraft was launched on 10 February 2020 and entered its Nominal Mission Phase in November 2021. The Polarimetric and Helioseismic Imager on the Solar Orbiter mission \citep[SO/PHI; see][]{solanki_2020} infers the photospheric magnetic field and line-of-sight (LoS) velocity from images of polarised light. It does this by sampling the \ion{Fe}{i}\,$6173$\,\AA\ absorption line at five wavelength positions and an additional point in the nearby continuum. Differential imaging is performed to acquire the Stokes $(I,Q,U,V)$ vector. \sophi\ has two telescopes: the High Resolution Telescope \citep[\hrt;][]{gandorfer2018high} and the Full Disc Telescope. In this paper only data from \hrt\ are discussed.

Solar Orbiter has a highly elliptic orbit with a perihelion as small as $0.28$\,au on some orbits. \sophi\ is the first magnetograph to move out of the Sun--Earth line. From 2025 on, with the help of Venus gravity assist manoeuvres, Solar Orbiter will reach heliolatitudes of $33^\circ$.

The Solar Dynamics Observatory \citep[SDO; see][]{pesnell2011solar} was launched on 11 February 2010 and orbits the Earth in a circular geosynchronous orbit with a $28\degree$ inclination. Like Solar Orbiter, SDO carries a magnetograph: the Helioseismic Magnetic Imager \citep[HMI;][]{scherrer2012helioseismic, schou2012design}. \hmi\ has been in regular science operations since 1 May 2010. Similar to \sophi, it samples the $6173$\,\AA\ \ion{Fe}{i} line at six points but at somewhat different wavelength positions.

 \begin{table*}
 \centering
      \caption[]{\hrt\ and SDO/\hmi\ instrument specifications.}
         \label{intro_table}
         \begin{tabular}{p{0.2\linewidth} R{6cm} R{6cm} }
             \hline
            \noalign{\smallskip}
            Specification  & \text{\hrt}  & \text{SDO/\hmi}\\
            \noalign{\smallskip}
            \hline
            \noalign{\smallskip}
            Working wavelength  & $6173$\,\AA   & $6173$\,\AA \\
            Wavelength positions & $-140,-70,0,70,140,$ $+$ or $-300$\,m\AA & $-172,-103,34,34,103,172$\,m\AA \\
            Field of view  & $0.28\degree\times0.28\degree$  & $0.57\degree\times0.57\degree$\\
            Aperture diameter & $140$\,mm & $140$\,mm    \\
            Spectral profile width & $106$\,m\AA & $76$\,m\AA \\
            Detector size & $2048\times2048$\,pixels  & $4096\times4096$\,pixels \\
            Plate scale & $0.5$\,\arcsec & $0.5$\,\arcsec\\
            Spatial resolution & $203$\,km (0.28\,au)\:-\:$725$\,km (1.0\,au) & $725$\,km\\
            
            \noalign{\smallskip}
            \hline
         \end{tabular}
  \end{table*}

The relevant technical details of \hrt\ and \hmi\ are shown in Table~\ref{intro_table}. As can be seen, \hrt\ and \hmi\ share some technical specifications: the same working wavelength, aperture diameter, and plate scale. It is important to know that, unlike \sophi, \hmi\ has two identical cameras.\ One is dedicated to the LoS observables -- the LoS magnetic field ($B_{\rm LOS}$) and the LoS velocity -- and is referred to as the `front camera'. The second camera, known as the `side camera', is used together with the front camera to capture the full Stokes vector, in order to retrieve the vector magnetic field.

With \sophi\ and \hmi\ now operating simultaneously, they provide excellent opportunities for new research that combines data from both instruments. For example, stereoscopy is now possible, allowing for simultaneous observations of the same feature on the solar surface from two different viewpoints. This can be used to investigate the Wilson depression of sunspots \citep[][]{romero} and test disambiguation techniques for the magnetic field azimuth \citep[][]{valori2022disambiguation,gherardo_23}. 
These and many other applications build on the premise that the two instruments provide very similar measurements of the magnetic vector. Here we test this assumption and compare the magnetic fields inferred by \hrt\ and \hmi\ and try to understand their similarities and differences. 

In Sect. \ref{sec:data} the data from both instruments used in this study and their properties are presented.  In Sect. \ref{sec:method} the detailed method for this comparison is given. The results of the comparison of the magnetic field data products from \hrt\ and \hmi\ are discussed in Sect. \ref{sec:comp}, and in Sect. \ref{sec:conc} we outline the conclusions reached from this work.


\section{Data}\label{sec:data}

The data used in this study are from 7 March 2022 (see Table~\ref{data_table}) and thus from around Solar Orbiter's inferior conjunction -- that is, when Solar Orbiter was on the Sun--Earth line -- which took place at 09:01:56 UTC (Coordinated Universal Time) on 7 March 2022. Solar Orbiter's elevation from the ecliptic plane was $2.949^\circ$ at inferior conjunction, and the effective angular separation between the two spacecraft during the observation period ranged from $3.006^\circ$ to $3.024^\circ$. During this time, Solar Orbiter was at a distance to the Sun of between 0.493\,au and 0.501\,au. On the photosphere, the nominal spatial resolution of \hrt\ at this distance is $363$\,km. In the common field of view (FoV) was a sunspot with negative polarity located at a heliocentric angle of $\mu=\cos\theta=0.87$ as seen by \hrt. 

\subsection{\hrt\ magnetic field}\label{subsec:hrt_data}

The \hrt\ data were collected to support a nanoflare and active region Solar Orbiter Observing Plan \citep[see][]{zouganelis20}. The raw data from this observation campaign were downlinked to Earth and processed using the on-ground data reduction and calibration pipeline \citep[][]{sinjan2022ground}. In addition, the data were processed to remove residual wavefront errors, which originate mostly from the telescope's entrance window. This was achieved using a point spread function (PSF) determined from phase diversity analysis \citep[][]{1992JOSAA...9.1072P,1994A&AS..107..243L}. Additionally, in the same processing step as the PSF deconvolution, a convolution with the instrument's theoretical Airy disc was performed. This produced data without optical aberrations, with increased contrast, and limited the noise that would otherwise be added by the deconvolution procedure. For further information regarding phase diversity  analysis and the \hrt\ PSF, we refer the reader to \cite{fatima-PD, fatima-PD2}.  To determine the magnetic field vector, the radiative transfer equation (RTE) was inverted with C-MILOS \citep[][]{suarez2007usefulness} in the full vector mode, which assumes a Milne-Eddington (ME) atmosphere and uses classical estimates (CE) as the initial conditions for the inversion \citep[][]{semel1967contribution, rees1979line, landi2004polarization}. For operational reasons, \hrt's Image Stabilisation System (ISS) was switched off. The \hrt\ LoS magnetograms used in this study were generated from the vector magnetic field obtained from the RTE inversion: 
$B_{\rm LOS} = B\,\cos\gamma$, where $B_{\rm LOS}$ is the LoS component of the magnetic field, $B$ is the field strength, and $\gamma$ is the angle of the field to the LOS. 

The data from this campaign were recorded with a $60$-second cadence. As shown in \citet{sinjan2022ground}, this mode results in quiet-Sun magnetograms with a noise of $8.3$\,G (with ISS on). Future investigations, using data planned to be gathered during Solar Orbiter's next inferior conjunction in March 2023, will attempt to quantify the impact of non-ISS operation on the comparison.

 \begin{table*}
 \centering
      \caption[]{Observation details of used \hrt\ and \hmi\ data.}
         \label{data_table}
         \begin{tabular}{p{0.15\linewidth}  R{6cm} | R{1cm} R{1cm} | R{1cm}  R{1cm} |}
            \hline
            \noalign{\smallskip}
             & \hrt & \multicolumn{4}{R{5.3cm}|}{SDO/\hmi} \\
            \noalign{\smallskip}
            \hline
            \noalign{\smallskip}
            Start time & 2022-03-07 00:00:09 \text{UTC} & \multicolumn{4}{R{5.3cm}|}{2022-03-07 00:00:00 \text{TAI}} \\
            End time & 2022-03-07 01:06:09 \text{UTC} & \multicolumn{4}{R{5.3cm}|}{2022-03-07 01:12:00 \text{TAI}}        \\
            Distance & $0.493 - 0.501$ au & \multicolumn{4}{R{5.3cm}|}{$0.992$ au} \\
            ISS mode & Off & \multicolumn{4}{R{5.3cm}|}{On}\\
            Processing & Ground & \multicolumn{4}{R{5.3cm}|}{Ground}\\
            RTE mode & C-MILOS: CE+RTE & \multicolumn{4}{R{5.3cm}|}{VFISV} \\
            \noalign{\smallskip}
             \hline
            \noalign{\smallskip}
             & Vector &  \multicolumn{2}{R{2.4cm}}{\text{Line of sight}} & \multicolumn{2}{|R{2.4cm}|}{\text{Vector}} \\
            \noalign{\smallskip}
            \hline
            \noalign{\smallskip}
            Cadence & $60$\,s & $45$\,s & $720$\,s & $90$\,s & $720$\,s   \\
            Number of datasets & $56$ & $56$ & $7$ & $38$ & $7$ \\
            \noalign{\smallskip}
            \hline
         \end{tabular}
  \end{table*}

\subsection{\hmi\ magnetic field}\label{subsec:hmi_data}

\hmi\ treats its LoS and vector data products separately, each having two options for observing cadence. For this comparison study, all four possible data products were compared with \hrt\ (see Table~\ref{data_table}). The vector data products were generated from the \hmi\ vector pipeline \citep[][]{hoeksema2014helioseismic}, while the LoS products were generated with an algorithm similar to that used by the Michelson Doppler Imager (MDI) on board the Solar \& Heliospheric Observatory, hereafter referred to as the MDI-like algorithm \citep[][]{couvidat2012line}. The \hmi\ LoS versus \hmi\ vector has been compared by \cite{hoeksema2014helioseismic}, who show that the MDI-like algorithm underestimates the field strength in the strong field regime ($\left|B_{\rm LOS}\right|>600$\,G) compared to the inversion result. The \hmi\ $45$-second and $720$-second LoS magnetograms have a noise level in the quiet Sun, near disc centre, of $7-9$\,G and $3-4$\,G, respectively \citep[][]{couvidat2016observables, liu2012comparison}. 

The $45$-second magnetograms are produced every $45$ seconds from an interpolation of Stokes $I + V$ and Stokes $I - V$ filtergrams from a $270$-second interval \citep[][]{liu2012comparison, couvidat2016observables}. Since 13 April 2016, the full Stokes vector has been captured at a $90$-second cadence and inverted to create the vector magnetic field data product. This cadence is achieved by combining images from both cameras \citep[][]{liu_baldner_bogart_duvall_hoeksama_norton_scherrer_schou_2016}. To produce the $720$-second vector data product, a weighted temporal average is made every $720$ seconds, combining $90$-second Stokes vector maps collected over a period of more than $20$ minutes and inverted using the very fast inversion of the Stokes vector (VFISV) ME code \citep[][]{hoeksema2014helioseismic, borrero2011vfisv}. In Sect.~\ref{sec:method} we describe the method by which we take the difference in interval and light travel time into account  to ensure co-temporal observations are compared.

\section{Method}\label{sec:method}
\begin{figure*}
  \centering
  \includegraphics[width=\linewidth, trim={2.5cm 2.5cm 3cm 3cm},clip]{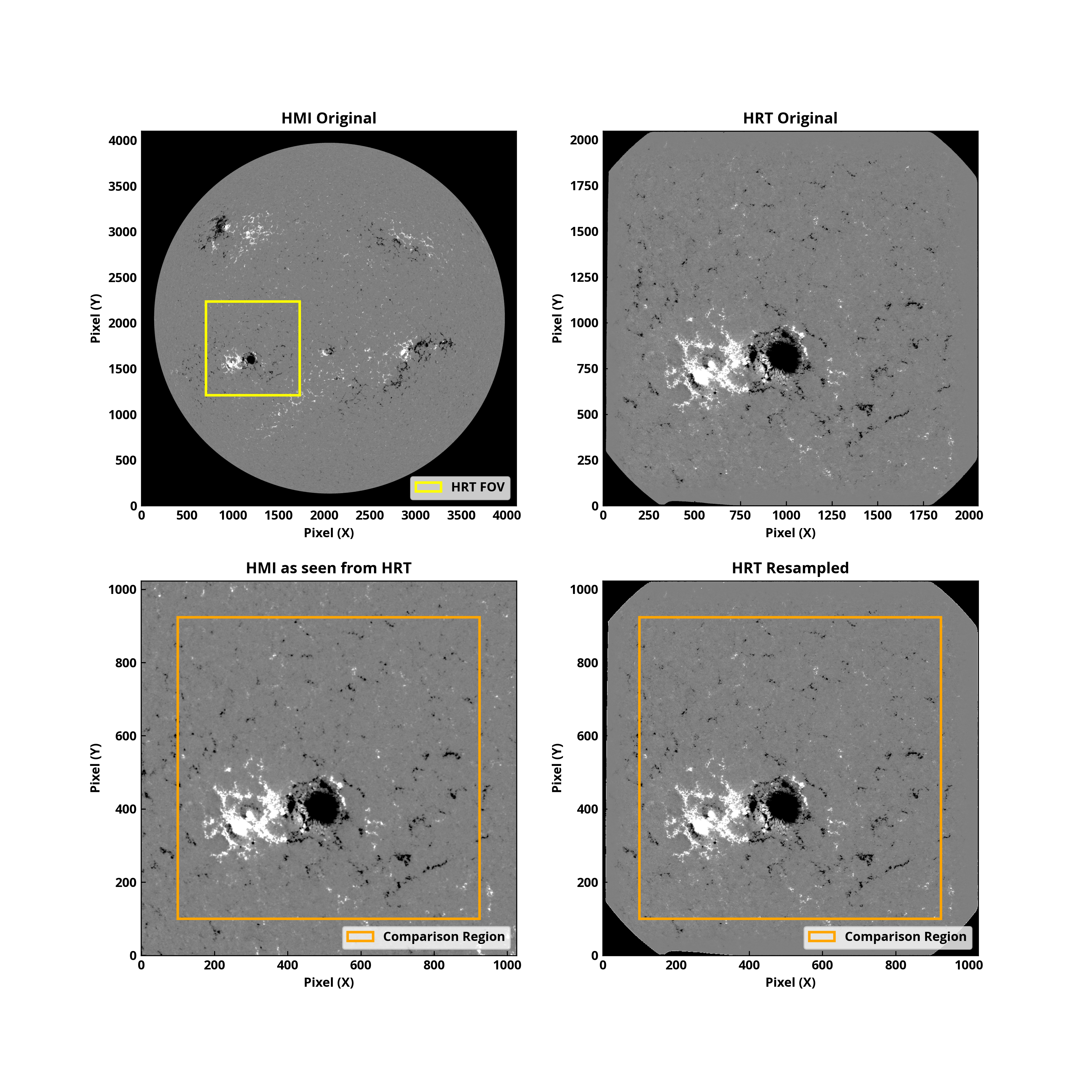}
  \caption{Magnetograms from \hmi\ and \hrt\ on 7 March 2022. Top left: \hmi\ $45$-second LoS magnetogram at 00:01:30 TAI, with the \hrt\ FoV shown in yellow. The pixels outside the solar disc are set to black for clarity. Top right: \hrt\ $60$-second magnetogram at 00:00:09 UTC. The pixels outside the field stop are set to black for clarity. Bottom left: Sub-region of the HMI magnetogram from the top-left panel, which has been re-projected to the \hrt\ frame of reference. Bottom right: \hrt\ magnetogram resampled to \hmi\ resolution. The orange square outlines the regions used for the comparison. All magnetograms are saturated at $\pm200\,$G.}
              \label{method_plot}%
\end{figure*}
    
\begin{figure*}
  \centering
  \includegraphics[width=\linewidth]{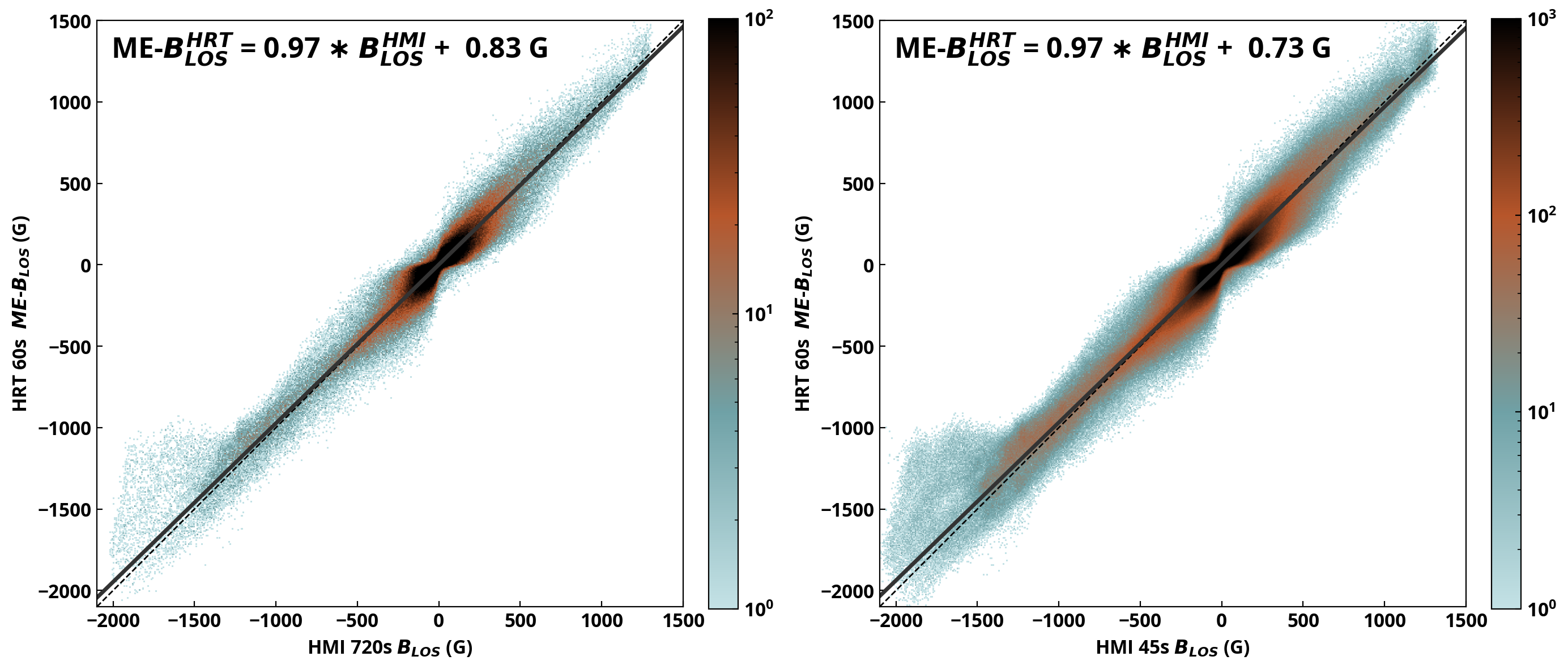}
  \caption{Scatter plot comparing pairs of \hrt\ $60$-second ME-$B_{\rm LOS}$ and \hmi\ $B_{\rm LOS}$. The log density of the pixels is shown and saturated at 100 (a) and 1000 (b) pixels per plotted point for clarity. The averaged linear fit (of \hmi\ vs \hrt\ and \hrt\ vs \hmi) is shown with the solid grey line, and a one-to-one correspondence is indicated by the dashed black line. Panel (a): Seven pairs with \hmi\ $720$-second magnetograms. Panel (b): $56$ pairs with \hmi\ $45$-second magnetograms. See the main text for a more detailed description.}
  \label{quad_plot}%
\end{figure*}

 We compared the magnetic field inferred by \hrt\ and \hmi\ on a pixel-to-pixel basis. The \hmi\ data were corrected for geometric distortion across the camera \citep[][]{hoeksema2014helioseismic}, and the \hrt\ data were corrected using a preliminary distortion model, derived from calibration data pre-launch. The method we now describe has been applied to each comparison of the individual data products. We provide here an example for one pair of LoS magnetograms: First a \hrt\ $60$-second magnetogram was selected and the closest \hmi\ $45$-second magnetogram in time was found (see the top panels in Fig. \ref{method_plot}). This was done by comparing the average time of the observations, taking into account the different distances of Solar Orbiter and SDO from the Sun, and hence the different light travel times, as well as the difference between TAI (International Atomic Time) and UTC time. Secondly, the sub-region of the \hmi\ FoV common to both telescopes, outlined in yellow in Fig.~\ref{method_plot}, was re-projected using the \citet{deforest2004re} algorithm onto the \hrt\ detector frame of reference using the World Coordinate System (WCS) information \citep[][]{thompson2006coordinate}. 

Next, the \hrt\ data were resampled using linear interpolation to match the factor of two lower spatial resolution of the \hmi\ data (\hrt\ was half the distance to the Sun at the time of observation). Applying boxcar binning or cubic interpolation makes no significant difference to the results of the comparison. As both \hrt\ and \hmi\ have the same aperture diameter, their PSFs are similar. However, by resampling \hrt\ we change the effective PSF. The impact of this effect is left for future studies. Residual rotation and translation perpendicular to the normal of the \hrt\ image plane were found using a log-polar transform \citep[cf. e.g.][]{sarvaija2009} and corrected. The result of such corrections is shown in the bottom panels of Fig.~\ref{method_plot}. These corrections are due to inaccuracies in the WCS information. This process was repeated for each \hrt\ magnetogram.

Finally, the maps were cropped by $100$ pixels at each side before the comparison was made, as outlined in orange in the lower panels of Fig.~\ref{method_plot}. This is because of the \hrt\ field stop, visible as the black region in Fig.~\ref{method_plot}, and because of the processing step to correct for residual wavefront errors. Within this procedure the image is apodised before the Fourier transform to ensure periodic boundaries, and the first $100$ pixels at each side were affected. These regions therefore had to be excluded from the comparison with \hmi.  

For comparison with the \hmi\ $720$-second data products, a single \hrt\ dataset, the one closest to the average time of the \hmi\ $720$-second dataset, was used. This comparison was performed for the LoS magnetic field component, $B_{\rm LOS}$, the magnetic field strength, $\left|\bm{B}\right|$, the inclination, $\gamma$, and the azimuth, $\phi$. Extra treatment was taken for the azimuth comparison: both \hmi\ and \hrt\ define the azimuth anti-clockwise from the positive direction of the $y$-axis \citep[][]{sinjan2022ground}. After the re-projection of HMI, care was taken to ensure that both datasets used the same definition of the azimuth by taking the roll angle of each spacecraft into account. 

\section{Comparison of \hrt\ and \hmi\ magnetic field observations}\label{sec:comp}
\subsection{Comparison of \hrt\ and \hmi\ LoS magnetograms}\label{subsec:blos}

We stress here for clarity that, when discussing the LoS magnetograms from \hmi, we refer to the LoS magnetic field derived using the MDI-like algorithm, referred to as $B_{\rm LOS}$. However, the magnetograms from \hrt\ presented here are the LoS component of the full vector magnetic field (determined by RTE inversion): we refer to this as ME-$B_{\rm LOS}$.

The scatter plot comparing the \hrt\ 60-second and \hmi\ 720-second magnetograms is shown in Fig.~\ref{quad_plot}a, where the logarithmic density of the points is indicated by the colour. This figure displays seven pairs of magnetograms; each of the \hrt\ 60-second magnetograms is recorded in the middle of the interval of time over which the \hmi\ 720-second magnetogram that it is compared with is recorded. The solid black line is a linear fit to the distribution, which is the average of two linear fits, one of \hmi\ versus \hrt\ and the other of \hrt\ versus \hmi. This averaging removes statistical bias. As indicated by the fit, there is an excellent agreement between the two telescopes, with a slope value of $0.97$ and an offset of $0.83$\,G. This offset could be an artefact of there being more very strong fields with negative polarity than with positive. The offset of the weak fields inferred by \hrt\ can be determined by histogram analysis: \cite{sinjan2022ground} demonstrate that the \hrt\ $B_{LOS}$ distribution in the quiet Sun is centred near zero with an offset of $-0.18$\,G. The Pearson correlation coefficient is $0.97$. The linear fit, absolute error on the slope and offset, and Pearson correlation coefficient (cc) are shown in Table~\ref{err_table} for all compared quantities presented in this paper. In the case of Fig.~\ref{quad_plot}, the errors on the slope and offset are negligible.

However, a difference is present for the strongest fields. We selected pixels where \hmi\ $720$-second $B_{\rm LOS}<-1300$\,G, the point at which  a large divergence between \hrt\ and \hmi\ appears. The mean difference between them is $+149\pm2$\,G relative to the (negative) \hmi\ values, which corresponds to $9$\% weaker LoS fields relative to \hmi. The error here denotes the standard error in the mean; the scatter ($1\sigma$) of the distribution of absolute differences is $197$\,G. The pixel selection threshold (\hmi\ $720$-second $B_{\rm LOS}<-1300$\,G) corresponds to pixels only in the leading sunspot in the FoV, where $81$\,\% are in the umbra and the remaining $19$\,\% in the penumbra. The umbra and penumbra classification was determined using $I_{c} < 0.55$ and $0.55 I_{c} < 0.95$ thresholds on the \hrt\ continuum intensity, $I_{c}$; these thresholds are the same as those used in \cite{dalda2017statistical}, where the magnetic field between \hmi\ and Hinode/SP is compared. It must also be noted that the distribution in Fig.~\ref{quad_plot} is not symmetric between fields of opposite polarity. This is because no strong fields above $1350$\,G were observed in \hmi\ in the common FoV, while \hrt\ infers fields of up to $1500\,$G. Under similar conditions, we expect the comparison between the two telescopes in the positive strong field regime to be similar to that observed in the negative strong field regime, with \hrt\ measuring lower LoS field components compared to \hmi.

The comparison with \hmi\ $45$-second magnetograms (Fig.~\ref{quad_plot}b), where $56$ pairs of data were compared, reveals very similar results. This was expected as the $45$-second and $720$-second HMI magnetograms are well inter-calibrated \citep{liu2012comparison}. For pixels where the \hmi\ $45$-second $B_{\rm LOS}<-1300$\,G, there is a similar mean difference of $+155.5\pm0.9$\,G relative to the (negative) \hmi\ values, which again corresponds to $9\%$ weaker LoS magnetic fields inferred by \hrt\ in this regime.

In both Fig.~\ref{quad_plot}a and b, all pixels are plotted, including those with signal below the noise. There is an hourglass shape around the origin present in both panels. This could be due to a mismatch in the alignment of the sets of magnetograms. As described in Sect.~\ref{sec:method}, we applied only a preliminary model to correct for geometric distortion in \hrt, which could explain inaccuracies in the alignment. 

There are several effects that could explain the difference between \hrt\ and \hmi\ for the strongest fields. Firstly, the two instruments use different methods to infer the LoS magnetic field: \hmi\ uses the MDI-like formula, while \hrt\ uses a radiative transfer code. Additionally, the two instruments sample the \ion{Fe}{i} line at different positions, and \hrt\ observes farther out in the continuum ($\pm300$\,m\AA\ from the line core vs $\pm172$\,m\AA\ for \hmi). For the strongest fields, the very large Zeeman splitting results in the two instruments capturing different information from the true Stokes signal, which is then interpreted by the inversion routines differently. A detailed investigation of these effects is beyond the scope of this paper. Furthermore, the spectral profile width is different: \hrt\ has a full width half maximum (FWHM) of $106$\,m\AA, while the FWHM of \hmi\ is\ $76$\,m\AA. There could also be a contribution from stray light, in particular for the pixels in the umbra, as neither \hmi\ nor \hrt\ are corrected for stray light in their standard data pipelines.

Finally, it is known that \hmi\ suffers from a 24-hour periodicity \citep[][]{liu2012comparison, hoeksema2014helioseismic, couvidat2016observables} in its magnetic field observables due to the SDO orbit. The velocity relative to the Sun oscillates by $\pm3.5$\,km/s on a 24-hour period, with further variation of hundreds of metres per second due to Earth's orbit. The SDO solar radial velocity for the data considered in this study started at $3.249\,$km/s and ended at $3.291\,$;m/s.  \citet{couvidat2016observables} show that the $B_{\rm LOS}$, calculated using the MDI-like algorithm, in the umbra depends quadratically on the magnitude of the velocity. A residual of between $+50$\,G and $+100$\,G was present when SDO had a radial velocity near $\pm3\,$km/s. This residual is the value once the long-term variations ($\geq 2$\,day) are removed. It explains approximately half of the observed difference in the strong signal regime.  It is plausible, although not certain, that, when combined with the effects from the different wavelength sampling, different inversion codes, and stray light, it explains the observed discrepancy.

\subsection{Comparison of \hrt\ and \hmi\ vector magnetic fields}

\begin{figure*}
  \centering
  \includegraphics[width=0.947\linewidth]{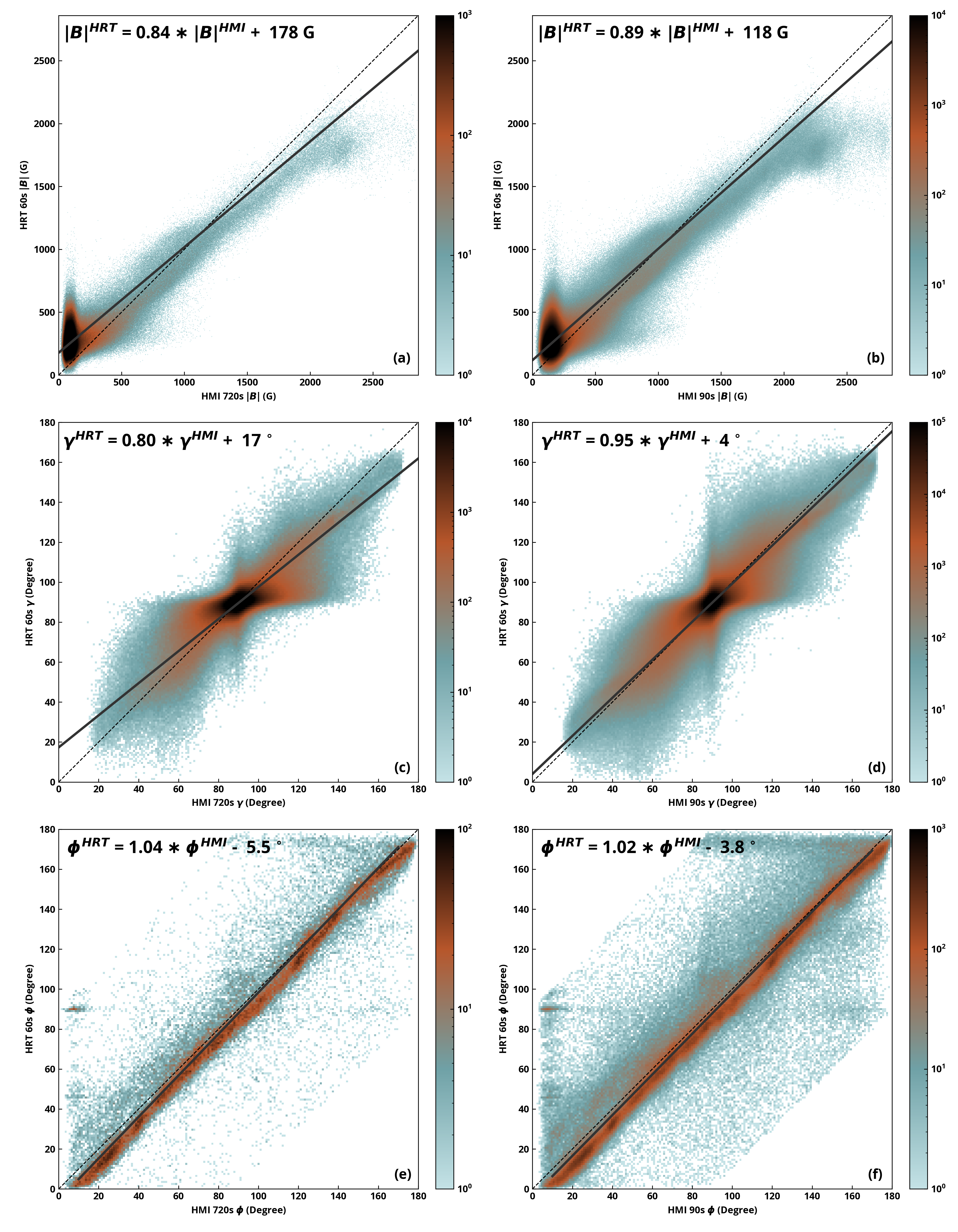}
  \caption{Scatter plots comparing \hrt\ and \hmi\ vector magnetic field maps. The first column compares inversion results from seven pairs of \hrt\ $60$-second and \hmi\ $720$-second datasets, while the second column does the same for $38$ pairs of \hrt\ $60$-second and \hmi\ $90$-second datasets. The log density of the pixels is given by the colour scale and is saturated for clarity. The averaged linear fit and $y=x$ are given by the solid grey and dashed black lines, respectively. Panels (a) and (b): Magnetic field strength. Panels (c) and (d): Magnetic field inclination (relative to the LoS). Panels (e) and (f): Magnetic field azimuth. Pixels where $|\phi_{\text{HMI}}-\phi_{\text{HRT}}| > 90\,\degree$ and $\mathbf{\left| B\right|}_{HRT} < 600\,$G are omitted and not included in the fit.}
              \label{vec_quad_plot}%
    \end{figure*}
    
Here we compare the \hrt\ and \hmi\ vector magnetic fields, both inferred by RTE inversions of the Stokes vector albeit using different inversion codes. We would like to highlight the $3^{\circ}$ angular separation between \hrt\ and \hmi, mentioned in Sect.~\ref{sec:data}. This has no impact on $\mathbf{\left| B\right|}$, and from a simple rotation test on \hrt\ data, we estimate that it does not significantly impact the magnetic field inclination or azimuth, except for producing an offset of a few degrees in the azimuth.

First we compared the magnetic field strengths, $\mathbf{\left| B\right|}$, as shown in the top row of Fig.~\ref{vec_quad_plot}. Both \hrt\ and \hmi\ assume a magnetic filling factor of unity for the RTE inversion, so the field strength is averaged over the pixel. Consequently, we do not distinguish between magnetic field strength and magnetic flux density, as is sometimes done in the literature. In Fig.~\ref{vec_quad_plot}a the comparison between the \hrt\ and \hmi\ $720$-second $\mathbf{\left| B\right|}$ is depicted, while in Fig.~\ref{vec_quad_plot}b the comparison with the \hmi\ $90$-second $\mathbf{\left| B\right|}$ is shown. The slope is $0.84$ and $0.89$ in Fig.~\ref{vec_quad_plot}a and b, respectively. The higher slope value for the $90$-second comparison is because the variance is more similar to that of the \hrt\ data than for the \hmi\ $720$-second data. The magnetic field strengths of the two instruments have a correlation coefficient of $0.85$ and $0.84$ for the $720$-second and $90$-second $\mathbf{\left| B\right|}^{HMI}$, respectively.

\begin{figure*}
  \centering
  \includegraphics[width=0.96\linewidth]{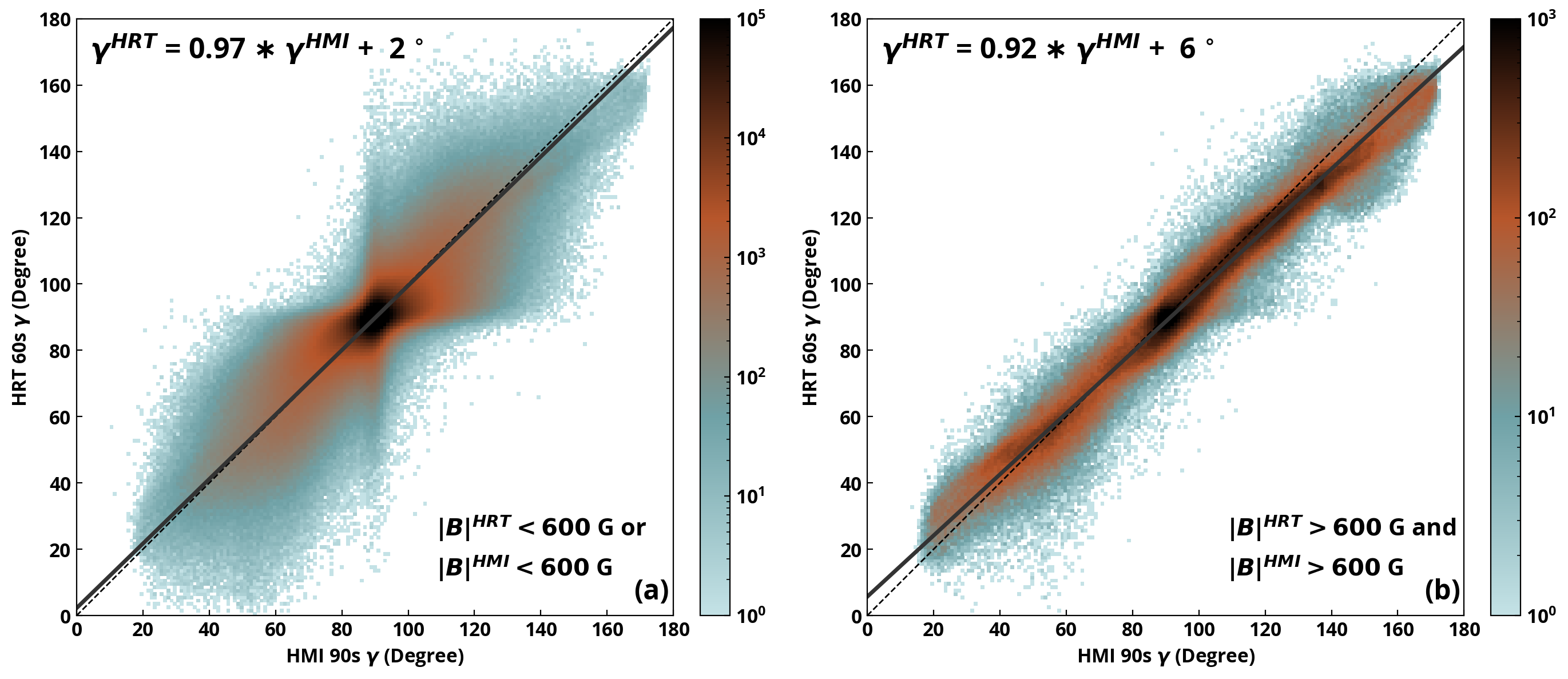}
  \caption{Scatter plots comparing \hrt\ $60$-second and \hmi\ $90$-second magnetic field inclination. Panel (a): Pixels where $\mathbf{\left| B\right|}_{\text{HRT}} < 600$\,G or $\mathbf{\left| B\right|}_{\text{HMI}} < 600$\,G. Panel (b): Pixels where $\mathbf{\left| B\right|}_{\text{HRT}} > 600$\,G and $\mathbf{\left| B\right|}_{\text{HMI}} > 600$\,G. The log density of the pixels is shown and is saturated for clarity. The averaged linear fit and $y=x$ are shown with the solid grey and dashed black line, respectively.}
              \label{binc_plot}%
    \end{figure*}

We observe here that in the weaker field regime, \hrt\ infers stronger fields. Following \cite{liu2012comparison}, we arbitrarily used a boundary value of $600$\,G to define the weak signal regime. In this regime there is a dense distribution of pixels, seen in both Fig.~\ref{vec_quad_plot}a and b, which we refer to as the `hot zone', that portrays a discrepancy between the two instruments. The offset is mainly due to this hot zone, with an offset of $178$\,G in Fig.~\ref{vec_quad_plot}a and a lower offset of $118$\,G in Fig.~\ref{vec_quad_plot}b. The difference in the offset perhaps reflects the noise difference between the $90$-second and $720$-second $\mathbf{\left| B\right|}^{HMI}$. The hot zone in Fig.~\ref{vec_quad_plot}b has a larger extent for \hmi\ compared to that in Fig.~\ref{vec_quad_plot}a, which may be due to the difference in noise level.
\cite{borrero2011inferring} have demonstrated that Stokes profiles with higher noise levels, when inverted, result in stronger but more inclined fields. We note the more horizontal dense field central patches in Fig.~\ref{vec_quad_plot}c, Fig. 3d, and Fig.~\ref{binc_plot}a. The higher noise level in \hrt\ compared to \hmi\ is due to the ISS non-operation and, crucially, the longer averaging time within the \hmi\ data. Furthermore, the deconvolution of part of the PSF also increased the noise of the \hrt\ data by $20$\% \citep[][]{fatima-PD2}. Therefore, the noise levels of the original Stokes vector in \hrt\ are $1.8\times10^{-3},\,2.2\times10^{-3}$, and $\,1.8\times10^{-3}$ for $Q/I_{c}$, $U/I_{c}$, and $V/I_{c}$, respectively, where $I_c$ denotes Stokes $I$ in the continuum. In comparison, the noise in the \hmi\ $720$-second Stokes vector is $9\times10^{-4}$ for $Q/I_{c}$, $U/I_{c}$,  and $V/I_{c}$, \citep[][]{couvidat2016observables}. The noise in the $90$-second Stokes vector, however, has not been quantified in the literature because this is a non-standard data product.

Now we turn to the strong signal regime in Fig.~\ref{vec_quad_plot}a and b. At approximately $\mathbf{\left|B\right|}\,>1300$\,G for both \hmi\ and \hrt,\, the distribution starts diverging from the $y=x$ line. For pixels where the fields in \hmi\ are stronger than this value, \hrt\ infers a lower field strength. The field strength threshold of $1300$\,G in \hmi\ and \hrt\ corresponds to pixels where $38.1$\,\% are in the umbra, $61.4$\,\% are in the penumbra, and $0.5$\,\% lie elsewhere. For fields stronger than $1300$\,G in \hrt\ or \hmi, the mean difference between them was $-247\pm1$\,G and $-246.8\pm0.4$\,G relative to the \hmi\, for the \hmi\ $720$-second and $90$-second comparisons, respectively ($\approx13$\% smaller relative to the \hmi\ values in both cases). The error on the mean is the standard error. The scatter ($1\sigma$) of the distribution of the differences is roughly $180$\,G in both cases, highlighting the large width of these distributions. While we cannot directly compare these mean differences to those presented in Sect.~\ref{subsec:blos}, because the strong magnetic field lines are not all along the LoS and we consider more pixels in the penumbra, we can still qualitatively deduce that we observe a larger separation between \hmi\ and \hrt\ for the magnetic field strength. Important to note is that in Fig.~\ref{quad_plot} we compare the ME-$B_{LOS}^{HRT}$, which was derived from the full vector, while the $B_{LOS}^{HMI}$ in Fig.~\ref{quad_plot} was calculated using the MDI-like formula. In Sect.~\ref{subsec:me_blos} the LoS components derived from the full vectors are compared.

The inclination of the magnetic vector, $\gamma$, relative to the LoS, as deduced from the two instruments, is compared in the second row of Fig.~\ref{vec_quad_plot}. The slope is $0.80$ and $0.95$ for the \hmi\ $720$-second and $90$-second comparisons, respectively. The correlation coefficient between \hrt\ and the \hmi\ $720$-second and $90$-second magnetic field inclination is $0.81$ and $0.85,$ respectively. It is clear that both instruments agree on the polarity of the magnetic field relatively well (there is a dearth of points in the upper-left and lower-right quadrants of Fig.~\ref{vec_quad_plot}c and d). We also note here that \hmi\ has a somewhat stronger tendency to infer inclinations close to $90\,^{\circ}$ (the vertical streak at $90\,^{\circ}$ is stronger than the horizontal one). The biggest difference between the inclinations inferred by the two instruments is, however, that \hrt\ data result in somewhat more horizontal fields (the slope of the solid black lines in Fig.~\ref{vec_quad_plot}c and d is less than unity). There is a closer agreement in Fig.~\ref{vec_quad_plot}d, with a slope of $0.95$, as the variance in the \hmi\ $90$-second data is closer to the \hrt\ variance. The offsets shown in both Fig.~\ref{vec_quad_plot}c and d are not relevant here as the point of symmetry lies at ($90^{\circ},90^{\circ}$). The averaged linear fit crosses ($90^{\circ},90^{\circ}$) with an offset of less than half a degree in both Fig.~\ref{vec_quad_plot}c and d. From the simple rotation test on \hrt\ data mentioned earlier, the $3^{\circ}$ angular separation between \hrt\ and \hmi\ could introduce an offset of $<1^{\circ}$. Furthermore, a small part of the scatter -- the distance of the points from the line of best fit -- is likely due to the $3^{\circ}$ difference in view direction.
 
In Fig.~\ref{binc_plot} we compare the inclination for the weak and strong field cases. In Fig.~\ref{binc_plot}a pixels are shown where $\mathbf{\left| B\right|}_{\text{HRT}} < 600$\,G or $\mathbf{\left| B\right|}_{\text{HMI}} < 600$\,G, while in Fig.~\ref{binc_plot}b pixels are shown where $\mathbf{\left| B\right|}_{\text{HRT}} > 600$\,G and $\mathbf{\left| B\right|}_{\text{HMI}} > 600$\,G. In Fig.~\ref{binc_plot}b the distribution of the points is much closer to the line of best fit, with a correlation coefficient of $0.98$, compared to a correlation coefficient of $0.80$ in Fig.~\ref{binc_plot}a. The slope in Fig.~\ref{binc_plot}b, however, is slightly lower than that in Fig.~\ref{binc_plot}a.

The comparison of the azimuth, $\phi$, is shown in the bottom row of Fig.~\ref{vec_quad_plot}. For this comparison, only pixels from and around the leading sunspot in the FoV, with $B_{\text{HRT}} > 600\,$G, were selected. Furthermore, for the linear fit, pixels where $|\phi_{\text{HMI}}-\phi_{\text{HRT}}| > 90\,\degree$ were not considered as they are affected by the intrinsic $180\,\degree$ ambiguity of the azimuth. Finally, the regions near $0\,\degree$ and $180\,\degree$ were excluded from the linear fits to avoid an artificial shift, as the end points were not periodic. There are strong correlation coefficients of $0.95$ and $0.94$ (\hmi\ $720$-second and $90$-second comparisons, respectively). One reason why there is a strong correlation is that the \hmi\ transverse magnetic field does not suffer from the $12-$ or $24$-hour periodicity due to the SDO orbit \citep[][]{hoeksema2014helioseismic}. As shown in Fig.~\ref{vec_quad_plot}e and f, the slope is $1.04$ and $1.02,$ respectively, implying that \hrt\ infers azimuth angles slightly larger than that of \hmi. There is also a negative, non-uniform offset of $-5.5\,^{\circ}$ in the $720$-second case, which is only $-3.8\,^{\circ}$ in the $90$-second case; this requires further investigation. The absolute errors on these offset values are $0.7$ and $1.7$, which are large relative to the computed offsets, as fewer points are considered relative to the other comparisons presented in this work. Were there an incorrect alignment of the $+y$ detector between \hrt\ and \hmi,\, which both define $\phi=0$, an offset between $\phi_{HMI}$ and $\phi_{HRT}$ would exist. To the best of our knowledge, we have aligned the $+y$ detector of both to solar north and thus rule this out as an origin of the observed offset. However, our rotation test also revealed that a rotation around axes orthogonal to the $+y$ detector axis could also result in an offset of $0^{\circ}$-$2^{\circ}$. Therefore, a part of the offset shown in Fig.~\ref{vec_quad_plot} could originate from the angular separation between \hrt\ and \hmi. In this test, the slope of the linear fit between the rotated and original \hrt,\ $\phi$, was $1.01$, a change of $1\%$, which is reflected in the slope error for the $\phi$ comparisons in Table~\ref{err_table}. Additionally, a small part of the scatter may be due to the $3^{\circ}$ angular separation.

Something that could explain the discrepancies seen in all three components of the magnetic vector is the different wavelength sampling and spectral resolution, as mentioned in Sect.~\ref{subsec:blos}. This, combined with the use of different inversion routines (VFISV applied to \hmi\ data and C-MILOS to \hrt\ data), is certain to result in differences between the two instruments. As mentioned in the discussion of the weak magnetic field strength regime, the difference in noise levels, in part due to longer \hmi \ integration times, is the reason for the different inferred fields. A non-perfect alignment of the data, as mentioned in Sect.~\ref{subsec:blos}, could also be a factor in explaining the noted difference.

\subsection{Comparison of \hrt\ and \hmi\ LoS components of the full vector magnetic field}\label{subsec:me_blos}
\begin{figure*}
  \centering
  \includegraphics[width=0.96\linewidth]{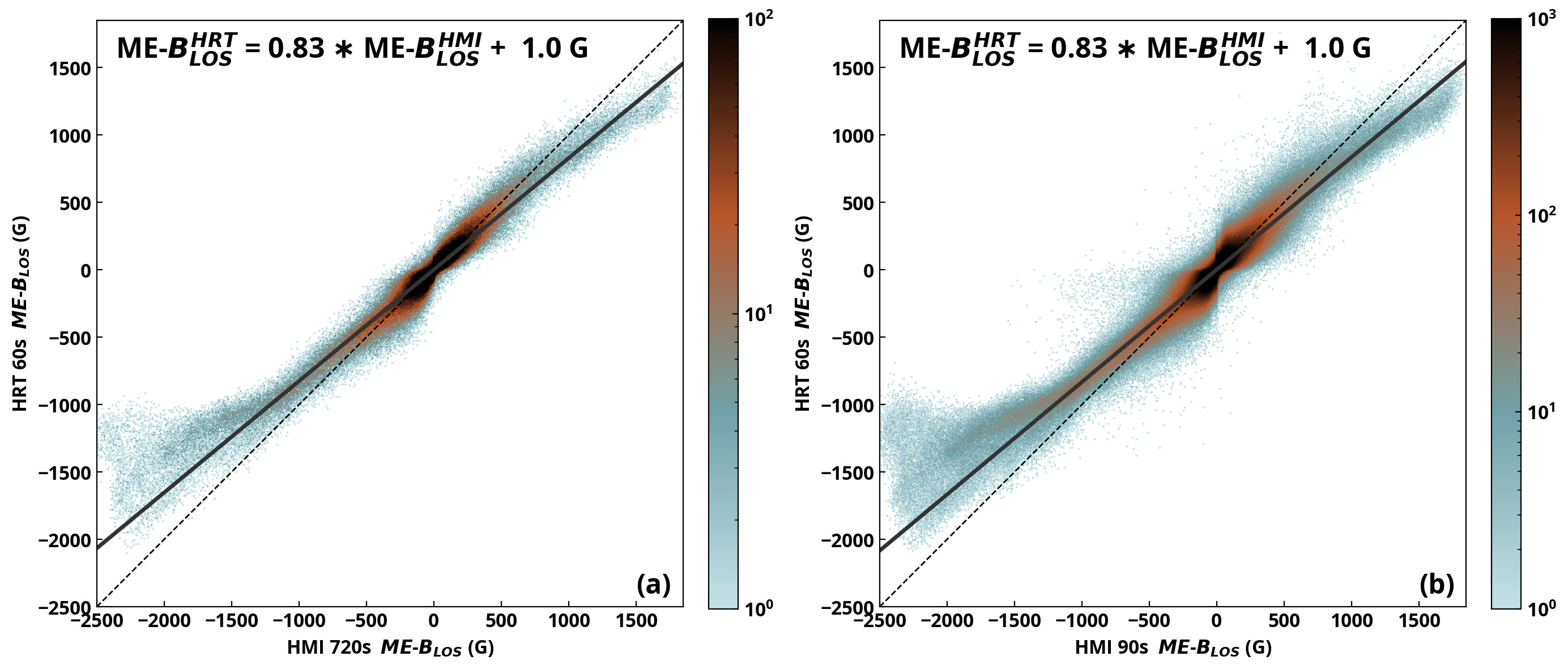}
  \caption{Scatter plots comparing the \hrt\ and \hmi\ LoS components of the full vector magnetic field. `ME' stands for Milne-Eddington and indicates that it is derived from RTE inversions. Panel (a): Comparison of inclinations from seven pairs of \hrt\ $60$-second and \hmi\ $720$-second data. Panel (b): Same, but for $38$ pairs of \hrt\ $60$-second and \hmi\ $90$-second data. The log density of the pixels is shown by the colour and is saturated at 1000 (panel a) and 100 (panel b) for clarity. The averaged linear fit and $y=x$ lines are plotted in solid grey and dashed black lines, respectively.}
              \label{me_blos_comparison}%
    \end{figure*}

\begin{table*}
 \centering
      \caption[]{Quantities compared, their linear fit, absolute errors on the slope and offset, and Pearson correlation coefficient (cc).}
        \label{err_table}
        \begin{tabular}{c|c|c|c|c}
        \hline
        \noalign{\smallskip}
        Quantities compared & Linear fit & Slope error & Offset error & Pearson cc \\
        \noalign{\smallskip}
        \hline
        \noalign{\smallskip}
        ME-$B_{LOS}^{HRT}$ $60$\,s vs $B_{LOS}^{HMI}$ $720$\,s & ME-$B_{LOS}^{HRT} = 0.97 \ast B_{LOS}^{HMI} + 0.83$\,G & $9\times10^{-5}$ & $0.01$ & 0.97 \\[5pt]
        ME-$B_{LOS}^{HRT}$ $60$\,s vs $B_{LOS}^{HMI}$ $45$\,s & ME-$B_{LOS}^{HRT} = 0.97 \ast B_{LOS}^{HMI} + 0.73$\,G & $3\times10^{-5}$ & $0.006$ & 0.97 \\[5pt]
        $\mathbf{\left|B\right|}^{HRT}$ $60$\,s vs $\mathbf{\left|B\right|}^{HMI}$ $720$\,s & $\mathbf{\left|B\right|}^{HRT} = 0.84 \ast \mathbf{\left|B\right|}^{HMI} + 178$\,G & $3\times10^{-4}$ & $0.02$ & $0.85$ \\[5pt]
        $\mathbf{\left|B\right|}^{HRT}$ $60$\,s vs $\mathbf{\left|B\right|}^{HMI}$ $90$\,s & $\mathbf{\left|B\right|}^{HRT} = 0.89 \ast \mathbf{\left|B\right|}^{HMI} + 118$\,G & $1\times10^{-4}$ & $0.01$ & $0.84$ \\[5pt]
        $\gamma^{HRT}$ $60$\,s vs $\gamma^{HMI}$ $720$\,s & $\gamma^{HRT} = 0.80 \ast \gamma^{HMI} + 17\,^{\circ}$ & $4\times10^{-4}$ & $0.01$ & $0.81$ \\[5pt]
        $\gamma^{HRT}$ $60$\,s vs $\gamma^{HMI}$ $90$\,s & $\gamma^{HRT} = 0.95 \ast \gamma^{HMI} + 4\,^{\circ}$ & $1\times10^{-4}$ & $0.004$ & $0.85$ \\[5pt]
        $\phi^{HRT}$ $60$\,s vs $\phi^{HMI}$ $720$\,s & $\phi^{HRT} = 1.04 \ast \phi^{HMI} - 5.5\,^{\circ}$ & $0.01$ & $0.7$ & $0.95$ \\[5pt]
        $\phi^{HRT}$ $60$\,s vs $\phi^{HMI}$ $90$\,s & $\phi^{HRT} = 1.02 \ast \phi^{HMI} - 3.8\,^{\circ}$ & $0.01$ & $1.7$ & $0.94$ \\[5pt]
        $\gamma^{HRT}$ $60$\,s vs $\gamma^{HMI}$ $90$\,s (weak-field) & $\gamma^{HRT} = 0.97 \ast \gamma^{HMI} + 2\,^{\circ}$ & $1\times10^{-4}$ & $0.006$ & $0.80$ \\[5pt]
        $\gamma^{HRT}$ $60$\,s vs $\gamma^{HMI}$ $90$\,s  (strong-field)& $\gamma^{HRT} = 0.92 \ast \gamma^{HMI} + 6\,^{\circ}$ & $2\times10^{-4}$ & $0.02$ & $0.98$ \\[5pt]
         ME-$B_{LOS}^{HRT}$ $60$\,s vs ME-$B_{LOS}^{HMI}$ $720$\,s & ME-$B_{LOS}^{HRT} = 0.83\, \ast \,$ME-$B_{LOS}^{HMI} + 1.0$\,G & $1\times10^{-4}$ & $0.01$ & $0.97$ \\[5pt]
        ME-$B_{LOS}^{HRT}$ $60$\,s vs ME-$B_{LOS}^{HMI}$ $90$\,s & ME-$B_{LOS}^{HRT} = 0.83\, \ast \,$ME-$B_{LOS}^{HMI} + 1.0$\,G & $5\times10^{-5}$ & $0.005$ & $0.95$ \\[5pt]
        \noalign{\smallskip}
        \hline
         \end{tabular}
  \end{table*}

We compare the LoS magnetograms from \hrt\ (from RTE inversions) with those inferred by \hmi\ (also from RTE inversions) in Fig.~\ref{me_blos_comparison}. The correlation coefficient is $0.97$ and $0.95$ for the $720$-second and $90$-second case, respectively, while the slope is $0.83$ for both. We detect here a systematic difference in the strong field regime, with \hrt\ inferring weaker LoS fields. \cite{hoeksema2014helioseismic} report that the \hmi\ MDI-like $B_{\rm LOS}$ underestimates the fields in comparison to the \hmi\ ME-$B_{\rm LOS}$. Therefore, as the \hrt\ ME-$B_{\rm LOS}$ agrees well with the \hmi\ MDI-like $B_{\rm LOS}$, as illustrated in Sect.~\ref{subsec:blos}, one expects to observe the same underestimation. We confirm this expectation here. Since the inclination is well correlated for strong fields (see Fig.~\ref{binc_plot}), we can determine that this observed difference is due to the overestimation of $\mathbf{\left| B\right|}$ by \hmi\ (or equally, the underestimation by \hrt). In comparison with Fig.~\ref{quad_plot} from Sect.~\ref{subsec:blos}, we can see that \hmi\ ME-$B_{\rm LOS}$ infers stronger LoS fields, up to $-2500\,$G and $1800$\,G, than those inferred with the MDI-like formula. Furthermore, the mean difference where \hmi\ ME-$B_{\rm LOS} < -1300$ is $486\pm2$\,G and $491\pm1$\,G for the $720$-second and $90$-second cases, respectively. These are roughly three times larger than those found in Sect.~\ref{subsec:blos}. The scatter ($1\sigma$) on these difference distributions is $239$\,G and $247$\,G, respectively.

Like the LoS magnetograms from \hmi, the LoS component of the vector magnetic field, the ME-$B_{\rm LOS}$ from \hmi,\ is also affected by the radial velocity of SDO. However, while the residual of the $B_{\rm LOS}$ calculated using the MDI-like algorithm varies quadratically with radial velocity, the residual of the \hmi\ vector LoS component varies linearly. At $+3$\,km/s, a residual of approximately $-30$\,G is determined, suggesting that \hmi\ may even be slightly underestimating the values compared to when SDO is at a radial velocity of $0$\,km/s \citep[][]{couvidat2016observables}. The effect from the radial velocity therefore cannot explain why \hmi\ infers a stronger field than \hrt\ in this comparison. 

\section{Conclusions}\label{sec:conc}

In this paper we have compared the magnetic fields inferred by \hrt\ and \hmi\ near the inferior conjunction of Solar Orbiter in March 2022. A comparison was made between the \hrt\ LoS component of the full vector magnetic field with both the \hmi\ $45$-second and $720$-second LoS magnetograms computed with the MDI-like algorithm. The \hrt\ ME-$B_{\rm LOS}$ and the \hmi\ $B_{\rm LOS}$ have a high correlation coefficient of $0.97$, a slope of $0.97,$ and an offset of less than $1$\,G. There is a difference, however, for the strongest fields ($B_{\rm LOS} < - 1300$\,G), where \hrt\ infers fields $9$\,\% smaller. These LoS fields correspond to regions in the leading sunspot in the umbra and penumbra only. There are too few points with $B_{LOS}>1300$\,G in the analysed dataset to determine if positive polarity fields recorded by the two instruments also display a difference. It is unclear what causes the difference at high field strengths. It could be that \hrt\ is saturated, or it could be due to the orbit-induced periodicity in \hmi\, as SDO was near its maximum radial velocity relative to the Sun at the time of co-observation. Other factors, such as the different wavelength sampling positions, inversion routines, and stray light, likely also contributed.

The vector magnetic fields inferred by \hrt\ and \hmi\ were also compared. 
Where $\mathbf{\left|B\right|}>1300$\,G, \hrt\ inferred field strengths $13$\,\% lower than \hmi, but with similar field inclination. This field strength threshold corresponded to regions almost exclusively in the umbra and penumbra in the active region in the common FoV. This is apparent in the comparison between the LoS component of the full vector magnetic field from both \hrt\ and \hmi. In the weak field regime ($\mathbf{\left|B\right|}<600$\,G), \hrt\ inferred stronger field strengths than \hmi. In this regime, the difference in field strength and inclination is mostly due to the difference in noise. The azimuth was compared by studying the large sunspot in the common FoV. It was shown to agree well, with a slope of $1.02-1.04$; however, there was a non-uniform, negative offset that requires further investigation.

The differences found between \hrt\ and \hmi, in both the LoS and vector magnetic fields, could be due to several factors. First of all, the two instruments sample different wavelength positions in the \ion{Fe}{i} absorption line and use different inversion routines to infer the vector magnetic fields. Secondly, there could be a non-perfect alignment in the magnetic field maps due to residual geometric distortion in the \hrt\ data. Additionally, neither the \hmi\ nor the \hrt\ data used in this study were corrected for stray light.
  
\begin{acknowledgements}
      This work was carried out in the framework of the International Max Planck Research School (IMPRS) for Solar System Science at the Max Planck Institute for Solar System Research (MPS). Solar Orbiter is a space mission of international collaboration between ESA and NASA, operated by ESA.  We are grateful to the ESA SOC and MOC teams for their support. The German contribution to SO/PHI is funded by the BMWi through DLR and by MPG central funds. The Spanish contribution is funded by AEI/MCIN/10.13039/501100011033/ (RTI2018-096886-C5, PID2021-125325OB-C5, PCI2022-135009-2) and ERDF “A way of making Europe”; “Center of Excellence Severo Ochoa” awards to IAA-CSIC (SEV-2017-0709, CEX2021-001131-S); and a Ramón y Cajal fellowship awarded to DOS. The French contribution is funded by CNES. The HMI data are courtesy of NASA/SDO and the HMI science team. This research used version 3.1.0 (10.5281/zenodo.5618440) of the SunPy open source software package \citep[][]{sunpy_community2020}. We are grateful to the anonymous referee for their constructive input.
\end{acknowledgements}

%
%

\bibliographystyle{aa}
\bibliography{bibfile.bib}






   
  



\end{document}